\documentstyle[aps,twocolumn,epsfig]{revtex}
\begin{document}
\newcommand{\ve}[1]{\mbox{\boldmath $#1$}}
\twocolumn[\hsize\textwidth\columnwidth\hsize
\csname@twocolumnfalse%
\endcsname

\draft

\title {Collisional relaxation in diffuse clouds of trapped bosons}
\author{G. M. Kavoulakis,$^1$ C. J. Pethick,$^1$ and H. Smith$^2$}
\date{\today}
\address{$^1$NORDITA, Blegdamsvej 17, DK-2100 Copenhagen \O, Denmark, \\
        $^2$\O rsted Laboratory, H. C. \O rsted Institute,
         Universitetsparken 5, DK-2100 Copenhagen \O, Denmark}
\maketitle

\begin{abstract}

    The damping of collective modes and the relaxation of temperature 
anisotropies in a trapped Bose gas is determined at temperatures above 
the Bose-Einstein condensation temperature in the collisionless 
regime.  We demonstrate for both cases how the effects of collisions 
may be treated perturbatively and calculate relaxation 
rates based on a variational principle.  Our results are compared with 
experiment and with previous theoretical calculations.

\end{abstract}

\vskip2pc]

\section{Introduction}

   In recent years a number of experiments have been carried out
to study relaxation processes in trapped Bose gases.
Such experiments may be divided into two broad categories, those
on collective oscillations and those on temperature
anisotropies. Relaxation rates of collective modes have been measured 
in rubidium vapors by Jin {\it et al.} \cite{jila1,jila2}, and in 
sodium ones by the MIT group \cite{mit,Ketnew}.  The relaxation of 
temperature anisotropies has been investigated by Monroe {\it et al.} 
\cite{Monroe} and Arndt {\it et al.} \cite{dalibard} in cesium, 
Newbury {\it et al.} \cite{Myatt} and Myatt \cite{myatt} in rubidium, 
and Davis {\it et al.} \cite{Davis} in sodium.

    In the experiments on the relaxation of
collective modes \cite{jila1,jila2,mit,Ketnew}, the atomic cloud is
excited at the frequency of the normal mode of interest by means of
a weak external perturbation. The amplitude of the oscillation is then
measured in the absence of the perturbation, and from that
the damping rate of the mode is extracted. In the
experiments on temperature relaxation, the atomic cloud
is either prepared in a nonequilibrium state with different
effective temperatures for motion along and perpendicular to the symmetry 
axis of the trap \cite{Monroe,dalibard,Myatt,myatt}, or it is driven out 
of equilibrium by displacing the trap center \cite{Davis}.

    Most experiments on relaxation have been carried out in the
collisionless regime, where the collision rate is small compared with
the frequencies of particle motion in the trap and also, in the case of 
collective oscillations, the oscillation frequency.  In addition the mean 
free path is large compared to the size of the cloud.  In this paper we 
focus on the collisionless limit, at temperatures above the 
condensation temperature, $T_c$, 
but our results also apply to the condensed phase 
at temperatures close to the transition temperature, where the thermal 
excitations constitute the dominant component of the gas.  A brief 
account of some of our results has been given previously \cite{kps}.

  Relaxation of temperature anisotropies in trapped clouds of
alkali-atom vapors has been studied theoretically both analytically
\cite{Myatt,myatt} and numerically \cite{Monroe,dalibard,Wu}.
Newbury {\it et al.} \cite{Myatt} and Myatt \cite{myatt}
used the Boltzmann equation with an ansatz for the form of the
non-equilibrium distribution function to obtain an analytic expression 
for the temperature relaxation rate and solved the problem numerically  
for large temperature anisotropies.  Monroe {\it et al.} \cite{Monroe} 
performed a Monte-Carlo simulation of temperature 
relaxation for atoms with an energy-independent scattering cross section, 
and Arndt {\it et al.} made a similar calculation for resonant 
scattering \cite{dalibard}.  Finally Wu and Foot solved the Boltzmann 
equation numerically and extracted the rate of temperature relaxation 
\cite{Wu}.

     The present paper is based on the semi-classical 
Boltzmann equation, from which we
derive an eigenvalue equation for the relaxation rate. Treating
the collisions perturbatively, we obtain general
expressions for the damping rate of the collective modes and the
relaxation rate for temperature anisotropies. We
include  effects of quantum degeneracy and
explore the role of energy dependence of
the scattering cross section.

    The paper is organized as follows: In Sec.\,II we  
develop the general formalism. 
We then examine in Sec.\,III the damping of modes with frequencies
equal to twice the characteristic frequencies of the trap
and calculate the damping rate for a general,
energy-dependent cross section.
The case of an energy-independent scattering cross section
is studied in detail, and analytical results
are obtained in the limit of Boltzmann statistics for a simple trial 
function.  We perform calculations with improved trial functions,
and find that the corrections are small, of the order of 1\%.
Our results are then compared with experiment.

    In Sec.\,IV we examine the relaxation of temperature anisotropies
and show that this problem is essentially the same as that of the 
damping of collective modes. Again we perform calculations with a
simple trial function and with a more general one. The improved 
trial function yields a reduction of the relaxation rate by about
7\%  for the two cases which we examine.
We then compare the results with data from experiment and from 
numerical simulations. Section V is a brief conclusion.

\section{General formalism}
 
When the energy of a particle is much greater than the spacing between energy 
levels the motion may be described semiclassically. In a trap, energy spacings 
are of order the oscillator quantum, $\hbar \omega_i$, where $\omega_i$ is 
the frequency associated with motion in the $i$ direction. Even when $T$ is as 
low as the transition temperature, the ratio $k_B T_c/\hbar \bar{\omega} =
[N/\zeta(3)]^{1/3}$ is very much greater that unity, where $N$ is
the total number of atoms in the trap, $\zeta(\alpha)$ is the Riemann 
zeta function, and $\bar{\omega}=(\omega_x \omega_y \omega_z)^{1/3}$,
and therefore the level spacing is small provided the anisotropy of the
trap is not extreme. Also we may neglect the influence of the mean
field due to interactions between atoms. This gives rise to a potential
of order $n({\bf r}) U_0$, where $n({\bf 
r})$ is the number density and $U_0 = 4 \pi \hbar^2 a/m$ is 
the effective two-particle interaction with $a$ being the 
scattering length for atom-atom collisions, and $m$ being the atomic 
mass.  Even when $T = T_c$, the energy $n U_0$ is typically only about 
$1\%$ of $k_BT$, and the molecular field
may therefore be neglected in the equation of motion of the particles.

   Since the motion can be described semiclassically  the state of the
system may be specified in terms of a particle distribution function 
$f({\bf p},{\bf r}, t)$,
where $\bf p$ is the momentum vector, $\bf r$ is the position vector, 
and $t$ is the time. This satisfies the Boltzmann equation,
\begin{eqnarray}
   \frac{\partial f}{\partial t} + \frac{\bf p}{m}\cdot {\ve \nabla} f  
- {\ve \nabla} V\cdot {\ve \nabla}_{\bf p} f =-I,
\label{BE}
\end{eqnarray}
where $V$ is the external potential, and $I$ is the collision
integral, which is a functional of $f({\bf r},{\bf p}, t)$. 
We write the distribution function as
$f= f^0 + \delta f$, where $f^0$ is the equilibrium distribution 
function and $\delta f$ is the deviation from equilibrium, and
linearizing Eq.\,(\ref{BE}) we find
\begin{eqnarray}
   \frac{\partial \, \delta f}{\partial t} + \frac{\bf p}{m}\cdot {\ve 
\nabla} \delta f  -
 {\ve \nabla} V\cdot {\ve \nabla}_{\bf p} \delta f =-I[\Phi],
\label{BEdeltaf}
\end{eqnarray}
where $\Phi$ is defined by $\delta f = f^0 (1+f^0) \Phi$.  

   We now consider
collisions.  The energy levels of a single particle moving in a harmonic
trap are discrete, and for a spherically symmetric trap there is strong
shell structure.  For the anisotropic traps used in experiments, the shell
structure is less pronounced, and the density of states will be made 
more uniform by interatomic interactions.  The ratio $n U_{0}/\hbar 
\bar{\omega}$ is of the order of unity for typical experimental conditions. 
Finally collisions will smear the density of states.  
As a consequence, we expect that under current experimental conditions the 
collision integral will be well approximated by the expression for a 
bulk system.  The linearized collision integral is thus
\begin{eqnarray}
   I[\Phi] = \int d\tau_1\int d\sigma \,|{\bf v}-{\bf v}_1|
  (\Phi+ \Phi_1- \Phi'- \Phi_1') \nonumber \\
  \times f^0 f^0_1 (1+{f^0}') (1+{f^0_1}').
\label{be}
\end{eqnarray}
Here the momenta of the incoming particles in a collision are denoted by 
$\bf p$ and ${\bf p}_1$, and those of the outgoing ones by ${\bf p}'$ and  
${\bf p}'_1$, and $d\tau=d{\bf p}/(2\pi\hbar)^3$. 
In the above expression we have introduced the differential cross
section $d\sigma$, which in general depends on the relative velocity
$u=|{\bf v}-{\bf v}_1|$ of the two incoming particles, as well as on the
angle between the relative velocities of the colliding particles before and 
after the collision. Since the atoms obey Bose-Einstein statistics, their
distribution function in equilibrium, $f^0({\bf p},{\bf r})$, is
\begin{eqnarray}
   f^0({\bf p},{\bf r}) = \frac 1 {e^{[p^2/2m+V-\mu]/k_BT} -1},
\label{equ}
\end{eqnarray}
where $\mu$ is the chemical potential. We shall consider harmonic
traps where the potential has the form
\begin{eqnarray}
  V= \frac 1 2 m(\omega_x^2 x^2+\omega_y^2 y^2 + \omega_z^2 z^2).
\label{potential}
\end{eqnarray}

    In the absence of collisions the motions of particles in the three
coordinate directions decouple, and particles execute simple harmonic 
motion for each of the coordinates.  Let us consider motion in the 
$z$ direction. In the absence of collisions ($I[\Phi] =0$) the general 
solution of the Boltzmann equation having time dependence 
$e^{-ni\omega_z t}$ has the form
\begin{eqnarray}
    \delta f = (p_z + i m \omega_z z)^n h(K_j),
\label{deltaf}
\end{eqnarray}
where $h$ is an arbitrary function, and $K_j$ are constants of the 
motion in the absence of collisions.
So, neglecting collisions, for $\delta f({\bf r},{\bf p}, t)=
h(K_j) (p_z + i m \omega_z z)^n e^{-ni\omega_z t}$, one has
\begin{eqnarray}
   \left( \frac{\partial }{\partial t} + n i \omega_z \right) \delta f 
  = 0.
\label{sol}
\end{eqnarray}
We now consider the effect of collisions.  Provided the collision rate 
is small compared with the oscillator frequency, collisions will not 
mix modes of the system in the absence of collisions with different 
values of $n$.  Consequently we expect the solution to be of the form
\begin{eqnarray}
   \delta f({\bf p},{\bf r}, t)=  h(K_j)
  (p_z + i m \omega_z z)^n e^{-ni\omega_z t -\Gamma t}.
\label{deltafc}
\end{eqnarray} 
This problem therefore amounts to doing degenerate perturbation theory
as in quantum mechanics, because we consider mixing of modes which have the 
same frequency in the absence of the perturbation, in this case collisions.  
Collisions will generally couple particle motions in different 
directions, but we begin by assuming that the collision rate is small 
compared with $|\omega_{z} - \omega_{x}|$ and $|\omega_{z} - \omega_{y}|$, 
in which case this coupling may be neglected.  It is 
important in traps with axial symmetry for modes in which particles 
move perpendicular to the symmetry axis, and we shall consider its 
effects later.  

    Substituting Eq.\,(\ref{deltafc}) into the Boltzmann equation
one finds
\begin{eqnarray}
   \left( \frac{\partial }{\partial t} + n i \omega_z \right) \delta f 
 = - \Gamma \, \delta f = -I[\Phi].
\label{soll}
\end{eqnarray}
Equivalently, Eq.\,(\ref{soll}) may be written as
\begin{eqnarray}
  \Gamma_i f^0 (1+f^0) \Phi_i = I[\Phi_i].
\label{eigenp}
\end{eqnarray}
The  problem thus reduces to solving the eigenvalue equation
(\ref{eigenp}) and it  can therefore be treated by conventional
perturbation theory, with the collision
integral being the perturbation.  We are interested in the mode 
with the smallest decay rate, since this is what is measured experimentally,
and therefore, rather than solving Eq.\,(\ref{eigenp}) 
directly, we use a variational method. A related variational method
is known to give highly 
accurate results in the calculation of transport coefficients 
\cite{Henrik}.  The most long-lived mode corresponds to the smallest 
value of $\Gamma_i$, which we denote by $\Gamma_0$.  Using 
Eq.\,(\ref{eigenp}) for any trial function $\Psi$, we multiply by 
$\Psi^*$ and integrate over the coordinates and momenta, and find that
\begin{eqnarray}
      \Gamma_0  \leq   \frac{\langle \Psi^*
    I[\Psi] \rangle}{\langle |\Psi|^2 f^0(1+f^0)  \rangle},
\label{assumption}
\end{eqnarray}
which is our basic variational expression for the damping rate.
Here the brackets denote integration over both coordinate and
momentum space. The numerator in the above equation expresses
essentially the entropy generation rate, whereas the denominator 
gives the energy in the mode \cite{Landau2}.

    It is worth pointing out the connection between 
Eq.\,(\ref{assumption}) and the usual variational expression for 
transport coefficients that applies in the static limit.  A transport 
coefficient such as the shear viscosity is determined by an 
inhomogeneous Boltzmann equation of the form
\begin{eqnarray}
    f^0 (1+f^0) X = I[\Xi],
\label{hydrod}
\end{eqnarray}
and the transport coefficient is proportional
to the quantity $\langle \Xi^* X f^0 (1+f^0) \rangle_{\bf p}$, where  
\begin{eqnarray}
  \langle \Xi^* X f^0 (1+f^0) \rangle_{\bf p} = \int d\tau \,
 f^0 (1+f^0) \Xi^*({\bf p}) X({\bf p}).
\label{scalarprod}
\end{eqnarray}
The above integration is over momentum space alone, in contrast to
the previous case [Eq.\,(\ref{assumption})], which involves integration 
over both coordinate and momentum space.
Then for any trial function $\Psi$ one finds \cite{Henrik} that
because the eigenvalues of the collision integral are never
negative, 
\begin{eqnarray}
    \langle \Xi^* X f^0 (1+f^0) \rangle_{\bf p}^{-1} \le \frac
   {\langle \Psi^* I[\Psi] \rangle_{\bf p}} 
  {\langle \Psi^* X f^0 (1+f^0) \rangle_{\bf p}^2}.
\label{teta2}
\end{eqnarray}

   Comparing Eqs.\,(\ref{assumption}) and (\ref{teta2}), we see that 
both are given by the same sort of variational expression and involve
the collision integral. The differences are
in the normalization condition which appears in the denominators of
the two expressions, and in the spatial integrations in 
Eq.\,(\ref{assumption}).

\section{Damping of modes}
\subsection{The damping rate}

   The inequality (\ref{assumption}) may be used to calculate
the damping rate of any mode in which particle motion in the $z$ 
direction is perturbed, but let us focus on a trap with uniaxial symmetry, 
$\omega_x = \omega_y$, and on modes with frequency $2 \omega_z$, which 
are among those of experimental interest.  In the same spirit as one 
adopts in calculating transport coefficients, we begin by making the 
simplest choice, Eq.\,(\ref{deltafc}) with $n=2$, and with $h$ put equal to 
unity.  Thus as our initial trial function $\Phi$ we use 
\begin{eqnarray}
   \Phi = (p_z + i m \omega_z z)^2.
\label{trfunction}
\end{eqnarray}
In Sec.\,III\,C we examine more general trial 
functions.  The function $\Phi$ contains terms of the form $z^2$, 
$zp_z$, and $p_z^2$.  It is clear from Eq.\,(\ref{be}) that the 
collision integral gives zero when operating on the first two of these 
terms because collisions conserve particle number and momentum; it 
also gives zero when operating on $p^2$ because of the conservation of 
the kinetic energy.  Therefore we may write
\begin{eqnarray}
   \langle (p_z - i m \omega_z z)^2 I[(p_z + i m \omega_z z)^2] 
\rangle=
\phantom{XXXX} \nonumber \\ \nonumber  \\
  = \langle p_z^2I[p_z^2] \rangle
 =\langle (p_z^2-p^2/3)I[p_z^2-p^2/3] \rangle .
\label{reduce}
\end{eqnarray}
The second of these forms follows as a consequence of kinetic
energy conservation, which allows us to substract any multiple of $p^2$ 
from $p_z^2$ in $I[p_z^2]$.  The choice made above introduces the 
function $p_z^2-p^2/3 $, which is orthogonal to the collision 
invariants, since $p_z^2-p^2/3 = (2/3) p^2 P_2(\cos \theta)$, where 
$P_l(x)$ is the Legendre polynomial of order $l$, and $\theta$ is the 
polar angle of the momentum vector.  Substitution of 
Eq.\,(\ref{reduce}) into Eq.\,(\ref{assumption}) leads to the result
\begin{eqnarray}
    \Gamma_0 \leq  \frac{ \langle
  (p_z^2-p^2/3)I[p_z^2-p^2/3] \rangle}
{\langle [p_z^2+(m\omega_z z)^2]^2 f^0(1+f^0)\rangle}.
\label{var2}
\end{eqnarray}
 
   This may be expressed in terms of the viscous
relaxation time $\tau_{\eta, {\rm var}}$ calculated within the
variational approximation. To see this, we start with the 
expression which connects $\tau_{\eta, {\rm var}}$ with the
viscosity $\eta$,
\begin{equation}
   \eta= \frac 1 m \langle |X|^2 f^0 (1+f^0) \rangle_{\bf p}
  \, \tau_{\eta,{\rm var}},
\label{viscc}
\end{equation}
where the trial function has been put equal to $X$, i.e., the driving 
term in the Boltzmann equation. The variational expression for the 
viscosity is
\begin{eqnarray}
   \eta^{-1} \le \eta_{\rm var}^{-1} = m
  \frac {\langle X^* I[X] \rangle_{\bf p}} 
 {\langle |X|^2 f^0 (1+f^0) \rangle_{\bf p}^2},
\label{ineqeta}
\end{eqnarray}
and therefore
we find with use of Eqs.\,(\ref{viscc}) and (\ref{ineqeta})
that Eq.\,(\ref{var2}) can be written as
\begin{equation}
     \Gamma_0 \leq \frac{1}{6} \frac {\int d{\bf r} 
    \,\tau_{\eta, {\rm var}}^{-1}
  \langle (p_z^2 - p^2/3)^2 f^0 (1+f^0) \rangle_{\bf p}}
 {\int d{\bf r} \, \langle (p_z^2 - p^2/3)^2 f^0 (1+f^0) \rangle_{\bf p}}.
\label{connection}
\end{equation}
The factor of 1/6 comes from the ratio of the integrals
$\langle (p_z^2 - p^2/3)^2 f^0 (1 + f^0) \rangle / 
\langle [p_z^2 + ( m \omega_z z)^2]^2 f^0 (1 + f^0) \rangle$ and reflects
the fact that collisions damp directly only
the part of the distribution function varying as $p_z^2-p^2/3$, while all
parts of the distribution contribute to the total energy associated with
the oscillation.
In the classical limit Eq.\,(\ref{connection}) takes the simple
form 
\begin{eqnarray}
    \Gamma_0 \leq \frac{1}{6}\int d{\bf r}\,    \frac{n({\bf
  r})}{\tau_{{\eta},\rm{var}}} \left/ \int d{\bf r} \,n({\bf r}) \right..
\label{sug}
\end{eqnarray}
 
   We return to Eq.\,(\ref{var2}) now, starting with the numerator 
on the right side of the inequality. This quantity includes integration
over real space as well as momentum space, and we write it as
\begin{eqnarray}
    \langle (p_z^2-p^2/3)I[p_z^2-p^2/3] \rangle =
  \int d{\bf r} \, I_c[z({\bf r})],
\label{nott}
\end{eqnarray}
where $z({\bf r}) = z(0) e^{-V/k_B T}$ is the fugacity, with $z(0) =
e^{\mu/k_B T}$ being its value at the center of the cloud.
$I_c[z({\bf r})]$ is given by
\begin{eqnarray}
  I_c[z({\bf r})] = \int d\tau \int d\tau_1\int d\sigma \,|{\bf 
v}-{\bf v}_1|
  \phantom{XXX}
 \nonumber \\
\times \Phi^* (\Phi+ \Phi_1- \Phi'- \Phi_1')
  f^0 f^0_1 (1+{f^0}') (1+{f^0_1}').
\label{be2}
\end{eqnarray}
In the above expression unprimed quantities refer to incoming
particles and primed ones to outgoing particles. Using the symmetry 
of the collision integral under interchange of incoming and outgoing
particle momenta, we may write Eq.\,(\ref{be2}) as
\begin{eqnarray}
    I_c[z({\bf r})] =
  \frac 1 4 \int d\tau \int d\tau_1\int d\sigma \, |{\bf v}-{\bf v}_1|
 \phantom{XXX}
\nonumber \\ \times
 |\Phi+ \Phi_1- \Phi'- \Phi_1'|^2
f^0 f^0_1 (1+{f^0}') (1+{f^0_1}').
\label{be3}
\end{eqnarray}
Because of the Galilean invariance of the collision process it is
convenient to 
use the center-of-mass momentum $\bf P$ and the relative momentum 
coordinates ${\bf p}_r$ and ${\bf p}'_r$ instead of the
momenta of the incoming and outgoing particles ${\bf p}, {\bf p}_1$,
and ${\bf p}', {\bf p}'_1$,
\begin{eqnarray}
    {\bf P} &=& {\bf p} + {\bf p}_1 = {\bf p}'+{\bf p}'_1;
 \phantom{XXXXXXXXX}
\nonumber \\
  {\bf p}_r &=& \frac 1 2 ({\bf p} - {\bf p}_1), \phantom{XX}
  {\bf p}'_r =  \frac 1 2 ({\bf p}' - {\bf p}'_1). 
\label{rel}
\end{eqnarray}
We introduce $\Delta[\Phi]$ by the definition
\begin{eqnarray}
  \Delta[\Phi] &=& \Phi + \Phi_1 - \Phi' - \Phi'_1,
\label{delta}
\end{eqnarray}
and for $\Phi = p_z^2-p^2/3$ one finds
\begin{eqnarray}
 \Delta[p_z^2-p^2/3] 
 &=& 2 ( p_{r,z}^2 - {p'_{r,z}}^{2}) \nonumber \\
  &=& 2 \left[ \left(p_{r,z}^2 - \frac 1 3 p_r^2 \right)
 - \left( {p'_{r,z}}^{2} - \frac 1 3 {p'_r}^2 \right) \right]
\nonumber \\
  &=& \frac 4 3 p_{r}^2
 \left[ P_2(\cos \theta_r) - P_2(\cos \theta'_r) \right],
\label{manip}
\end{eqnarray}
where $\theta_r$ and $\theta'_r$ are the polar angles of the vectors
${\bf p}_r$ and ${\bf p}'_r$ with respect to the $z$ axis. We write 
the element of cross section $d \sigma$ in Eq.\,(\ref{be2}) as
$(d \sigma/ d \Omega) d \Omega$, where $d \Omega$ is the element
of solid angle for the vector ${\bf p}'_r$. Since we are considering
scattering of indistinguishable bosons, the final states corresponding to
${\bf p}'_r$ and $-{\bf p}'_r$ are identical, and we perform the summation
over final states by summing over all $\Omega$ and then dividing by 2
to avoid double counting. Using the
addition theorem for Legendre polynomials and their orthogonality 
property, we find from Eq.\,(\ref{manip}) that the
numerator of Eq.\,(\ref{var2}) is
\begin{eqnarray}
    \langle (p_z^2-p^2/3)I[p_z^2-p^2/3] \rangle =
   \frac {16} {45 m} \int d{\bf r} \int d\tau \int d\tau_1
\nonumber \\ \times \frac 1 2 \int d\Omega \,
  \left[ 1 - P_2(\cos \chi) \right] \frac {d \sigma} {d \Omega}  \, 
 p_r^5 \, f^0 f^0_1 (1+{f^0}') (1+{f^0_1}'),
\label{nott2}
\end{eqnarray}
where $\chi$ is the angle between ${\bf p}_r$ and ${\bf p}'_r$. The product
of the Bose functions in Eq.\,(\ref{nott2}) depends on the angles 
between $\bf P$ and the vectors ${\bf p}_r$ and ${\bf p}'_r$, which 
implies that it is impossible to factorize the collision integral, 
except for the case of Boltzmann statistics.  In order to perform the 
angular integrations we express $\sin^2 \chi$ in terms of $\theta_r$ 
and $\theta'_r$, using the addition formula
\begin{eqnarray}
   \cos \chi = \cos \theta_r \cos \theta'_r +
 \sin \theta_r \sin \theta'_r \cos(\phi_r - \phi'_r),
\label{formula}
\end{eqnarray}
where $\phi_r$ and $\phi'_r$ are the azimuthal angles of the vectors
${\bf p}_r$ and ${\bf p}'_r$. Equation (\ref{formula}) then gives
\begin{eqnarray}
   \int_0^{2\pi} d(\phi_r-\phi'_r) \sin^2 \chi =
 \pi (1 + y^2 + y'^2 - 3 y^2 y'^2),
\label{formulaa}
\end{eqnarray}
with $y = \cos \theta_r$ and $y' = \cos \theta'_r$.

    The differential cross section $d\sigma/d\Omega$ depends in general 
on $y$ and $y'$ and in principle we may consider any form of 
$d\sigma/d\Omega$.  At the low energies of interest it is, however, a 
good approximation to assume that the scattering is purely $s$ 
wave \cite{Monroe,dalibard,Myatt,Dav}.  When the cross section is 
energy-independent, which is a very good approxiation for rubidium and 
sodium atoms, $d\sigma/d\Omega = 4 a^2$. The factor of 4 reflects
the fact that for identical bosons the scattering amplitude is twice 
that for distinguishable particles. For cesium atoms 
there is a bound state close to zero energy and the scattering is 
resonant. If the resonant state were at zero energy, the $s$ wave
phase shift $\delta_0$ would be $\pi$ (modulo $2 \pi$) and therefore
the cross section at low energy would be given by the unitarity limit
for $s$ waves, $d\sigma/d\Omega = 4/k^2$. More generally, when the
resonant state is close to, but not exactly at zero energy, one
may take the leading term in the effective range expansion 
\cite{Landau} and write $\tan \delta_0 = k a$. Since 
$d\sigma/d\Omega = (4 / k^2) \sin^2 \delta_0$, one then finds
\begin{eqnarray}
   \frac {d\sigma} {d\Omega} = \frac {4 a^2} {1 + (ka_{\rm 
sc})^2},
\label{deff}
\end{eqnarray}
where ${\bf k} = {\bf p}_r /\hbar$ is the wavevector
corresponding to ${\bf p}_r$. The differential
cross section  thus depends on the energy of the colliding
particles. For $s$ wave scattering the integral of the $P_2$ term
in Eq.\,(\ref{nott2}) vanishes, and the integrand on the right side 
in this equation
involves the total momentum-dependent
cross section $\sigma(u)$, which is defined as
\begin{eqnarray}
     \sigma(u) &=& \frac 1 2 \int d\Omega \, \frac {d \sigma} {d \Omega},
\label{momdeps}
\end{eqnarray}
where for the relative velocity of the incoming particles, ${\bf u}$,
${\bf u}=2 {\bf p}_r/m$. For a
differential cross section of the form of Eq.\,(\ref{deff})
the total cross section is
\begin{eqnarray}
  \sigma(u) = \frac {8 \pi a^2} {1 + (ka)^2}.
\label{momdeptsc}
\end{eqnarray}

   We turn now to the denominator of Eq.\,(\ref{var2}) and define $D$ as
\begin{eqnarray}
   D &=& \int d{\bf r}
 \int d\tau \, \left[ p_z^2 + (m \omega_z z)^2 \right]^2
 f^0 (1+f^0)  \nonumber \\
&=& 2 \int d{\bf r} \int d\tau \, \left[ p_z^4 +
p_z^2 (m \omega_z z)^2 \right]
 f^0 (1+f^0),
\label{denomin}
\end{eqnarray}
where the latter result follows from the fact that $f^0$ is a
function of $p_z^2 + (m \omega_z z)^2$, which is symmetric under
interchange of $p_z$ and $m \omega_z z$. The integral 
in Eq.\,(\ref{denomin}) is over 
a six-dimensional hyperspace with coordinates
${\bf p}$ and $m \omega {\bf r}$, and the ``angular'' and radial integrals
decouple. With the use of the equation $f^0 (1+f^0) = - k_B T \,
\partial f^0/\partial E$, with $E$ being the particle energy,
one finds
\begin{eqnarray}
   D &=& \frac 8 {15} \,
  \int d{\bf r} \int d\tau \, p^4 f^0 (1+f^0)
\nonumber \\ &=& \frac {8} {3} \, m k_B T \,
 \int d{\bf r} \int d\tau \, p^2 f^0
\nonumber \\ &=& \frac {2^{15/2} \pi} 3 (m k_B T)^{7/2}
 \int d{\bf r} \, \bar{D}({\bf r}),
\label{denomin2}
\end{eqnarray}
where 
\begin{eqnarray}
  \bar{D}({\bf r})& = &\int_0^{\infty} \frac {x^4 \, dx} {z^{-1}({\bf r})
 e^{x^2} - 1} \nonumber \\
 & = & \frac {3 \sqrt \pi} 8 g_{5/2}[z({\bf r})],
\label{corr00}
\end{eqnarray}
with $g_{n}(z) = \sum_{l=1}^{\infty} z^l/l^n$ being the familiar Bose 
integral.

     From Eqs.\,(\ref{nott2}), (\ref{momdeps}), (\ref{momdeptsc}) and
(\ref{denomin2}) we obtain
\begin{eqnarray}
  \Gamma_0 \le
   \frac {4} {15 m^2 k_B T} \int d{\bf r} \int d\tau \int d\tau_1
  \phantom{XXXXXXXXX} \nonumber \\ \times p_r^5 \, \sigma(u)
 f^0 f^0_1 (1+{f^0}') (1+{f^0_1}') \left/
\int d{\bf r} \int d\tau \, p^2 f^0 \right. .
\label{nott3}
\end{eqnarray}
Our final variational inequality (\ref{nott3}) can be written in the form
\begin{eqnarray}
   \Gamma_0 \leq \frac  1 {10 \sqrt 2} \,
 \left( \frac {m k_B T} {2 \pi \hbar^2} \right)^{3/2} \, \sigma_0 \, 
\bar{v}
\phantom{XXXXXXXXX}
\nonumber \\ \times
 \int \bar{I_c} [z({\bf r})] \, d{\bf r}
 \left/  \int d{\bf r} \, \bar{D}({\bf r})  \right. ,
\label{corr0}
\end{eqnarray}
which is one of our main results. Using Eq.\,(\ref{corr00}) we perform 
the spatial integration in the denominator of Eq.\,(\ref{corr0}) and find
\begin{eqnarray}
   \Gamma_0 \leq \frac  1 {30 \pi^{7/2} \sqrt 2}  \,
 \left( \frac {m \bar{\omega}} {\hbar} \right)^{3} 
  \frac {\sigma_0 \, \bar{v}} {g_4[z(0)]}
   \int \bar{I_c} [z({\bf r})] \, d{\bf r}.
\label{corr0rec}
\end{eqnarray}
In the above equations $\sigma_0 = 8 \pi a^2$
is the scattering cross section at zero energy, and 
\begin{eqnarray}
    \bar{v} = (8 k_B T/\pi m)^{1/2}
\label{defthv}
\end{eqnarray}
is the mean thermal velocity.
The quantity $\bar{I_c}$ is proportional to $I_c$ and it is
given by 
\begin{eqnarray}
    \bar{I_c} = \int_0^{\infty} dx_r \int_0^{\infty} dx_0
   \int_0^1 dy \int_0^1 dy' \, x_0^2 x_r^7 \, \bar{\sigma}(x_r,t)
\nonumber \\
\times  F(x_r,x_0,y,y')
 \, (1 + y^2 + y'^2 - 3 y^2 y'^2).
\label{Ic}
\end{eqnarray}
In the above expression 
we have introduced $\bar{\sigma}(x_r,t)$ which is given by the ratio
$\sigma(u)/\sigma_0$ and it is a function of $u = p_r a/\hbar$.
We have also defined $x_r=p_r/(m k_B T)^{1/2}$ and $t=T/T_0$, with $k_B T_0
= \hbar^2/ m a^2$, which implies that $u = x_r t^{1/2}$. Also 
$x_0$ is given by $x_0 = P/(4 m k_B T)^{1/2}$. Finally we have
introduced the product $F = f^0 f^0_1 (1+{f^0}') (1+{f^0_1}')$, which may
be written in the useful form
\begin{eqnarray}
   F = \frac {z^2 e^{-(x_0^2+x_r^2)} }
  {(1- z e^{-x_1^2}) (1- z e^{-x_2^2})
 (1- z e^{-x_3^2})(1- z e^{-x_4^2})},
\label{fuf}
\end{eqnarray}
where $x_i=p_i/(2m k_B T)^{1/2}$ with $p_i$ being the momentum of
particle $i$. The indices $i=1$ and 2 refer to incoming particles,
and $i=3$ and 4 to outgoing ones. The variables $x_i$
are explicitly given in terms of $x_r, x_0, y$, and $y'$ by
\begin{eqnarray}
   x_1^2 = \frac 1 2 (x_0^2 + 2 x_0 x_r y+ x_r^2),
  x_2^2 = \frac 1 2 (x_0^2 - 2 x_0 x_r y+ x_r^2),
\nonumber \\
 x_3^2 = \frac 1 2 (x_0^2 + 2 x_0 x_r y'+ x_r^2),
x_4^2 = \frac 1 2 (x_0^2 - 2 x_0 x_r y' + x_r^2).
\nonumber
\label{ksidef}
\end{eqnarray}

   Taking the classical limit of Eq.\,(\ref{corr0}), we find that
the integrand in the numerator is proportional to
$z^2({\bf r})$ and therefore to the square of the density,
as expected. The integrand in the spatial integration in the denominator
is linear in the density in this limit. In addition, in the classical 
limit the integrations over the center of mass and the relative
momenta decouple, and Eq.\,(\ref{corr0}) may be written
in the form
\begin{eqnarray}
     \Gamma_0 \leq \frac{\sqrt \pi} {30 \sqrt 2}
    \, \sigma_0 \bar{v}
   \left[ \int d{\bf r}\,  n^2_{\rm cl}({\bf r})  \left/
  \int d{\bf r} \,n_{\rm cl}({\bf r}) \right] \right.
 \phantom{XX} \nonumber \\ \times
\int_0^{\infty} dx_r \, x_r^7 e^{-x_r^2} \, \bar{\sigma}(x_r,t) \left/
\int_0^{\infty} dx \, x^4 e^{-x^2} \right.,
\label{newform}
\end{eqnarray}
where $n_{\rm cl}({\bf r}) = n_{\rm cl}(0) e^{-V/k_B T}$ is the density in
the classical limit, with $n_{\rm cl}(0)$ being the 
density of the atoms at the center of the cloud, where it is a maximium.

   To express Eq.(\ref{newform}) in a more physical form
it is convenient to define an average density weighted by the particle 
number, or
\begin{eqnarray}
     n_{\rm av} 
  = \int n_{\rm cl}^2({\bf r})
   \, d{\bf r} \left/ \int n_{\rm cl}({\bf r}) \, d{\bf r} =n_{\rm 
  cl}(0)/2^{3/2} \right. ,
\label{avedens}
\end{eqnarray}
where the last equality holds for a harmonic potential.
The result Eq.\,(\ref{newform}) may be written in a form that
makes it physically more clear,
\begin{eqnarray}
   \Gamma_0 &\leq& \frac{1}{12} \overline{\sigma(u) u}
  \int d{\bf r}\,    n^2_{\rm cl}({\bf r})  \left/
  \int d{\bf r} \,n_{\rm cl}({\bf r}) \right. \nonumber \\
 &\leq& \frac 1 {24 \sqrt 2} n_{\rm cl}(0) \overline{\sigma(u) u},
\label{classical}
\end{eqnarray}
where
\begin{equation}
  \overline{\sigma(u)u}  =\frac{\int_0^{\infty}du u^7\exp(-mu^2/4k_BT)
 \sigma(u)}{\int_0^{\infty}du
 u^6\exp(-mu^2/4k_BT)}.
\label{genn}
\end{equation}
Using Eq.\,(\ref{classical}) we find that classically for
energy-independent $s$-wave scattering,
\begin{eqnarray}
   \Gamma_0 \leq
   \frac 1 {15} n_{\rm cl}(0) \sigma_0 \bar{v} \equiv \Gamma_{0,{\rm cl}}.
\label{classical2}
\end{eqnarray}

   We turn now to the effects of quantum degeneracy. 
For energy-independent scattering and for $T \gg T_c$ one may derive
an analytical expression for the
damping, as shown in Appendix A, and the result is
\begin{eqnarray}
    \Gamma_0 \leq \frac{1}{15} \, n_{\rm cl}(0)
  \sigma_0 \bar{v}
 \left[ 1 + \frac 3 {16} \zeta(3) \left( \frac {T_c} T \right)^3
\right],
\label{corrnew}
\end{eqnarray}
where $n_{\rm cl}(0) = N  \bar{\omega}^3
[m /2 \pi k_B T]^{3/2}$.
The coefficient $3 \zeta(3)/16$ is approximately
equal to 0.23.
The dashed line in Fig.\,1 shows the result of the expansion given
by Eq.\,(\ref{corrnew}), as a function of $T_c/T$.
Under other conditions
the integrals in Eq.\,(\ref{corr0}) must be calculated numerically. 
The relaxation rate can be written as
\begin{eqnarray}
  \frac {\Gamma_0} {\Gamma_{0,{\rm cl}}} = G(T_c/T).
\label{simple}
\end{eqnarray}
The solid line in Fig.\,1 shows the function $G(T_c/T)$
for $s$ wave, energy-independent scattering. 
The value of $\Gamma_0$ at $T_c$ is $\approx 1.66\,
\Gamma_{0,{\rm cl}}$. The assumption of an
energy-independent cross section is a rather good approximation for
Rb and Na atoms, but later we shall consider the case of an
energy-dependent cross section, which is relevant for Cs.
\begin{figure}
\begin{center}
\epsfig{file=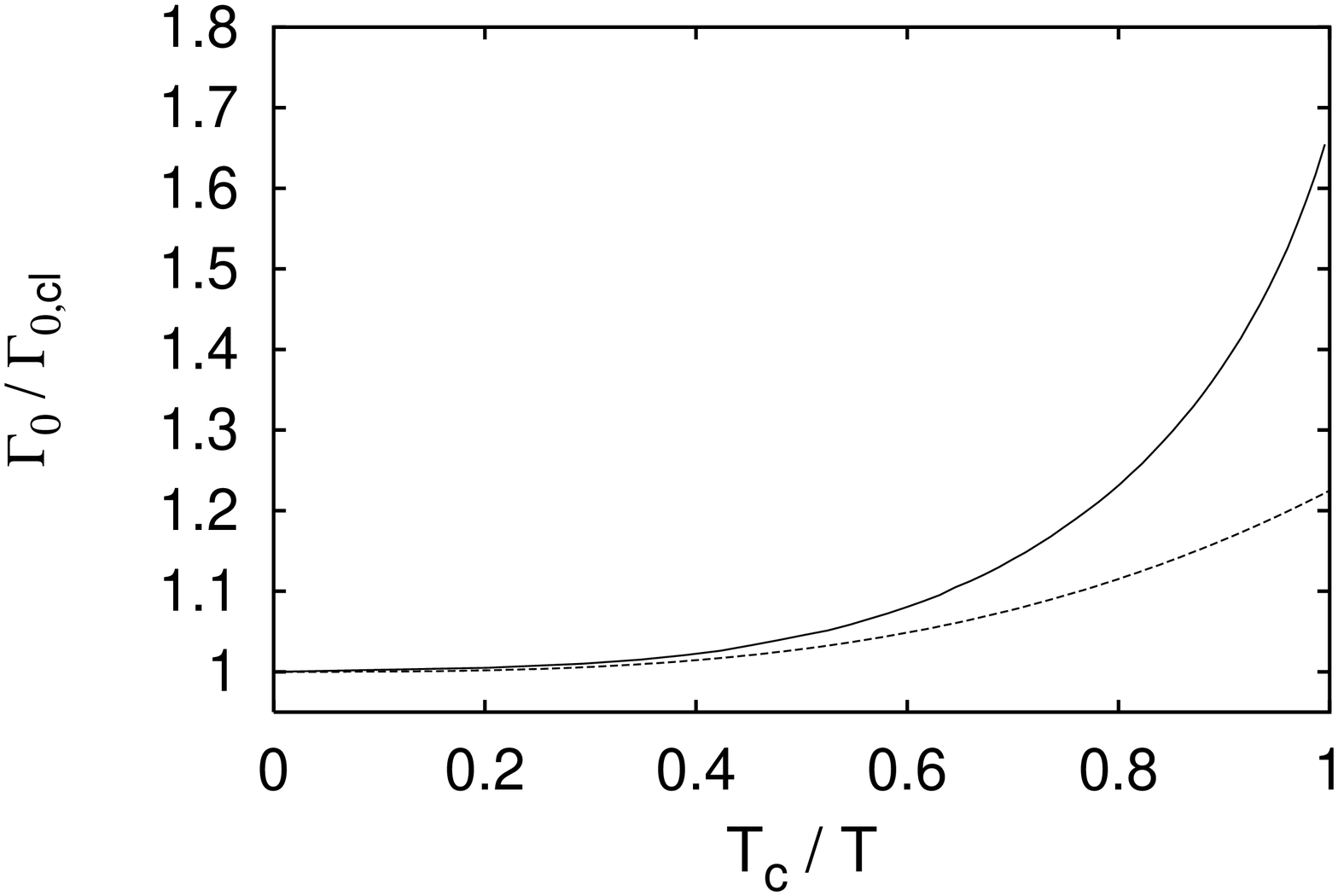,width=7.0cm,height=\linewidth,angle=-90}
\begin{caption}
{Solid line: The relaxation rate $\Gamma_0$ given by Eq.\,(\ref{corr0})
for energy-independent scattering, as a function of $T_c/T$,
normalized to the classical value, $\Gamma_{0,{\rm cl}}$.
Dashed line: The result of the analytic expansion given by
Eq.\,(\ref{corrnew})
also normalized to $\Gamma_{0,{\rm cl}}$.}
\end{caption}
\end{center}
\label{FIG1}
\end{figure}

\subsection{Traps with degenerate modes}

  Up to now we have studied oscillations along one of the axes of the 
cloud, taken to be the $z$ axis, and we have assumed 
that the mode is non-degenerate: $\omega_z \neq \omega_x$, and 
$\omega_z \neq \omega_y$.  There are experimental situations, however, 
where the damping of oscillations perpendicular to the axis of the 
trap has been measured \cite{jila1,jila2}.  
Since $\omega_x = \omega_y$, collisions couple the motion in the $x$ and $y$
directions. Because the collision integral is invariant under 
rotations of the spatial coordinates about the $z$ direction, the
appropriate solutions of the Boltzmann equation may be characterized
by the ``magnetic" quantum number, $m_z$, specifying the symmetry
under rotations about the $z$ axis. From the trial functions
$(p_x + i m \omega_x x)^2$ and $(p_y + i m \omega_y y)^2$ for motion
in the $x$ and $y$ directions one can construct two sorts of
trial function 
\begin{eqnarray}
   \Phi(m_z = 0) = (p_x + i m \omega_x x )^2 + (p_y + i m \omega_y y)^2,
\label{mz0}
\end{eqnarray}
and
\begin{eqnarray}
   \Phi(|m_z| = 2) = (p_x + i m \omega_x x )^2 - (p_y + i m \omega_y y)^2.
\label{mz2}
\end{eqnarray}
For $m_z=0$ the only part of this trial function that contributes to the  
collision integral is $p_x^2 + p_y^2$, and since $p_x^2 + p_y^2 = p^2 
- p_z^2$, the collision integral gives the same result as in the case 
studied in Sect.\,III A.  On the other hand with this choice of 
trial function the denominator is equal to $2D$, where $D$ is given by 
Eq.\,(\ref{denomin}), because the energy in the mode is twice as large.  This 
is easily seen by calculating $|\Phi(m_z=0)|^2$ and performing the angular 
integrations, which yield
\begin{eqnarray}
   \int d{\bf r} \int d\tau \, |\Phi(m_z = 0)|^2 f^0 (1+f^0) =
\phantom{XXXXXX}
\nonumber \\
 = 2 \int d{\bf r}
 \int d\tau \, \left[ p_x^2 + (m \omega_x x)^2 \right]^2
 f^0 (1+f^0) 
\nonumber \\
 = \frac {16} {15} \int d{\bf r} \int d\tau \, p^4 f^0 (1+f^0),
\label{mz0comp}
\end{eqnarray}
since the integral of the cross terms vanishes.
The denominator in the present problem is thus twice as large as the
one obtained in Eq.\,(\ref{denomin2}), and one finds
\begin{eqnarray}
  \Gamma(m_z=0) = \frac 1 2 \, \Gamma_0.
\label{half}
\end{eqnarray}
This result is independent of the degree of quantum degeneracy.

  Turning to the trial function with $|m_z|=2$, Eq.\,(\ref{mz2}),
we see that it is a superposition of terms with $m_z = \pm 2$. 
Since the $m_z = +2$ and $m_z = -2$ modes are degenerate, this 
mode has the same damping.
The only part of the trial function that contributes to the 
collision integral is $p_x^2-p_y^2$. Making use of the expressions
\begin{eqnarray}
   p_x^2-p_y^2 =(8 \pi/15)^{1/2} p^2 (Y_2^2+Y_2^{-2})
\label{mz4}
\end{eqnarray}
and
\begin{eqnarray}
   p_x^2+p_y^2 -2p^2/3= (2/3)^{1/2}(8 \pi /15)^{1/2} p^2 Y_2^0,
\label{mz6}
\end{eqnarray}
where $Y_l^m$ are the spherical harmonics, we find for the damping
rate $\Gamma(|m_z|=2)$ of the $|m_z|=2$ mode that
\begin{eqnarray}
   \Gamma_0(|m_z|=2) &=& 3 \,  \Gamma_0(m_z=0) \nonumber \\
                   &=& \frac 3 2 \, \Gamma_0,
\label{mz8}
\end{eqnarray}
since the denominator for $\Phi(|m_z| = 2)$ is the same as for $\Phi(m_z 
= 0)$. The above ratios are also independent of the quantum degeneracy.

\subsection{Improved estimates}

    One can improve the estimates of mode damping by using more general 
trial functions.  We start with Eq.\,(\ref{deltaf}), which gives the 
most general form of $\delta f$.  In the calculation presented in 
Sec.\,III A we assumed that the function $h(K_j)=1$, or
$\Phi_0 = (p_z + i m \omega_z z)^2$.
More general
trial functions can be used, and one systematic method for generating 
these is to expand $h$ in terms of complete sets of functions of the 
variables $K_j$.  For the problem under consideration the conserved 
quantities are the energies $E_i=p_i^2/2m + m\omega_i^2 r_i^2/2$ for 
each of the coordinate axes.  For a trap with an axis of symmetry the 
angular momentum about this axis is conserved, but this will not 
be a relevant variable, since in the experiments there 
was no preferred direction of rotation about the axis.  The number of 
possible functions of $E_i$ is vast, and we shall not perform a 
general search.  Rather we shall look for improved functions that 
depend only on the total energy.  Also for simplicity we consider 
the classical limit only. We expand $h$ in terms of a set of orthogonal 
polynomials, and a convenient choice is the Sonine polynomials 
introduced in the theory of transport coefficients in classical gases 
\cite{Burnett}.  Because of the form of the normalizing factors in the 
denominator in Eq.\,(\ref{assumption}) the appropriate orthogonality 
condition for the polynomials is
\begin{eqnarray} 
    \langle  h_i h_j E^2  f^0  \rangle  = \delta_{i,j},
\end{eqnarray}
where the factor $E^2$ reflects the presence of the prefactor 
$p_z^2 +(m\omega_z z)^2$ in the trial function, and $\delta_{i,j}$
is the Kronecker delta function.
The functions $h_i$ are thus proportional to the Sonine polynomials $S_4^i$ 
\cite{landau2}. We remark that the normalization we have chosen, which is 
convenient for our purposes, differs from the one used for Sonine 
polynomials. The simplest example is $h_0=1$, and the next one
is $h_1=\sqrt{5}(1-E/5kT)$ and thus
\begin{eqnarray}
  \Phi_1 = \sqrt 5 \, (p_z +i m \omega_z z)^2 (1 - E/5 k_B T).
\label{measure4}
\end{eqnarray}
As our improved trial function we take a linear combination of $h_0$ and 
$h_1$, which is equivalent to $c+E$, where $c$ is a constant.
We write the improved trial function as
\begin{eqnarray}
   \Phi_{\rm imp} &=& (1 - \alpha^2)^{1/2} \Phi_0 + \alpha \Phi_1,
\label{improved}
\end{eqnarray}
where $\alpha$ is a variational parameter. We proceed by calculating
the damping rate $\Gamma_{0,\rm imp}$ corresponding to this trial function
and minimizing it with respect to $\alpha$.
In terms of the quantities
\begin{eqnarray}
   \gamma_{i,j} = \langle \Phi_i^* I[\Phi_j] \rangle,
\label{gammaij}
\end{eqnarray}
we obtain
\begin{eqnarray}
   \Gamma_{0,\rm imp} =
  \frac 1 D  \left[(1-\alpha^2) \gamma_{0,0} + 2 \alpha 
(1-\alpha^2)^{1/2}
 \gamma_{1,0} \right. \nonumber \\ \left. + \alpha^2 \gamma_{1,1} 
\right].
\label{gamma0imp}
\end{eqnarray}
For the calculation of the quantities $\gamma_{1,0}$ and $\gamma_{1,1}$,
we notice that the only difficult part is the one that involves the 
combination
$(p_z + i m \omega_z z)^2 E/5 k_B T$. One can easily see that due to
the conservation laws, only the parts $(p^2/2m + V) p_z^2 + i \omega_z
z p_z p^2$ give non-vanishing terms, in addition to the terms we 
considered
earlier with $\Phi_0$ only. We find that in terms of the coordinates 
${\bf p}_r$, ${\bf p}'_r$, and $\bf P$, the combination of 
deviation functions (\ref{delta}) is given by
\begin{eqnarray}
    \Delta[\Phi_{\rm imp}] =
     2 \, (p_{r,z}^2 -{p'_{r,z}}^2)
   \left[ (1 - \alpha^2)^{1/2} + \alpha \sqrt 5
  \right. \nonumber \\ \left.
 - \frac {\alpha \sqrt 5} 5 \, \frac V {k_B T}
 + \frac {\alpha \sqrt 5} {10} \, \frac 1 { m k_B T} \left( p^2 +  
  \frac {P^2}  4
  \right) \right]
\nonumber \\ - \frac {\alpha \, 2 \sqrt 5} 5 \, \frac 1 {k_B T}
  \left( \frac {P_z} {2m} + i \omega_z z \right) 
 (p_{r,z} {\bf P} \cdot {\bf p}_r - p'_{r,z} {\bf P} \cdot {\bf p}'_r),
\label{expan}
\end{eqnarray}
which reproduces Eq.\,(\ref{manip}) for $\alpha=0$.
We calculate the matrix elements and find that 
$\gamma_{1,0}/\gamma_{0,0}=
-\sqrt 5 /20$, and $\gamma_{1,1}/\gamma_{0,0}=199/80$. Identifying the
ratio $\gamma_{0,0} /D$ as the quantity $\Gamma_0$ which we calculated
earlier, we write Eq.\,(\ref{gamma0imp}) as
\begin{eqnarray}
   \Gamma_{0,\rm imp} (\alpha)&=&\Gamma_0
  \left[ 1-\alpha^2 +2 \alpha (1 - \alpha^2)^{1/2} \frac
 {\gamma_{1,0}}{\gamma_{0,0}}
+ \alpha^2 \frac {\gamma_{1,1}} {\gamma_{0,0}} \right]
\nonumber \\
 &=&\Gamma_0 \left[ 1- \frac {\sqrt 5} {10} \alpha (1 - 
\alpha^2)^{1/2}
 + \frac {119} {80} \alpha^2 \right].
\label{finvaralpha}
\end{eqnarray}
Minimizing the above expression with respect to $\alpha$,
we find that $\alpha \approx 0.073$; the minimum is $\approx 0.992 \, 
\Gamma_0$
and therefore the correction is of the order of 0.8\%. Since $\alpha$
is very small, it is a good approximation to expand 
Eq.\,(\ref{finvaralpha}) in powers of $\alpha$
and write $\Gamma_{0,\rm imp}$ in the
form of a perturbative expansion as in quantum mechanics:
\begin{eqnarray}
   \Gamma_{0,\rm imp} = \Gamma_0 \,
  \left[ 1 - \frac {\gamma_{1,0}^2} {\gamma_{0,0}
 (\gamma_{1,1} - \gamma_{0,0})} \right].
\label{gamma0impp}
\end{eqnarray}
The correction term $1-\gamma_{1,0}^2/\gamma_{0,0}(\gamma_{1,1} - 
\gamma_{0,0})$
is equal to $118/119 \approx 0.992$. Our final result is therefore
\begin{eqnarray}
    \Gamma_{0,\rm imp} = \frac {118} {1785} 
  n_{\rm cl}(0) \sigma_0 \bar{v}.
\label{gamma0impp1}
\end{eqnarray}
This provides evidence for the result of the earlier calculations
being very close to the actual value.

\subsection{Comparison with experiment}

  The MIT group has reported two measurements of the
damping rate of oscillations along the axis of a cigar-shaped cloud.
In the first one \cite{mit} $N = 5 \times 10^7$ atoms \cite{private}
were observed to oscillate at a frequency of $\approx 35$ Hz at $T=2\,
T_c$ and the damping time was measured to be about $80$ ms.  The
frequencies of the trapping potential, which was axially symmetric,
were $\nu_z \approx 19$ Hz along the axis of the trap,
and $\nu_{\perp} \approx 250$ Hz perpendicular to it.
Using these numbers we find that classically the central density is
$\approx 2.9 \times 10^{13}$ cm$^{-3}$ and assuming that $a =
27.5$ \AA \cite{Dav} for Na atoms, we find from Eqs.\,(\ref{simple})
and (\ref{gamma0impp1})
that the damping time in the collisionless regime is $\approx 46$ ms.
However, the system is neither in the collisionless nor in the
hydrodynamic regime. The oscillation frequency is $\approx 35$ Hz,
which is intermediate between the one in the collisionless limit
$\nu_C = 2 \nu_z \approx 38$ Hz, and the one in the hydrodynamic limit
$\nu_H \approx \sqrt{12/5} \, \nu_z \approx 1.55 \, \nu_z\approx 29$
Hz \cite{hy}.  We can use the simple interpolation formula given in
Ref.\,\cite{KPS} to calculate the corrected damping time.  According
to this formula, the damping time is approximately $(\nu_C
-\nu_H)/(\nu -\nu_H)$ times the value we found earlier in the
collisionless limit, which gives a damping time of $\approx 70$ ms, in
rather good agreement with the experimental value of $\approx 80$ ms.
As shown in Refs.\,\cite{Usama}, a detailed study of the frequency
and the damping of the modes in the regime intermediate between
the hydrodynamic and the collisionless limits shows that for the
modes studied experimentally the interpolation formula we introduced
in Ref.\,\cite{KPS} is a very good approximation.

   In the other experiment performed by the MIT group, the number
of atoms was $N \approx 8 \times 10^7$ at $T \approx T_c$ and the
frequencies of the trapping potential were $\nu_{\perp}
\approx 230$ Hz, and $\nu_z \approx 17$ Hz.
The oscillation frequency was measured to be $\nu \approx 1.75 \, \nu_z$
and the damping time to be $\approx 50$ ms. The peak density calculated
classically is $n_{\rm cl}(0) \approx 6.3 \times 10^{13}$ cm$^{-3}$
and our estimate Eqs.\,(\ref{simple}) and (\ref{gamma0impp1})
for the damping time for this experiment
in the collisionless limit is $\approx 18$ ms at $T=T_c$. Using again the
interpolation formula from Ref.\,\cite{KPS} we find that the corrected
damping time is $\approx 41$ ms, which is also in reasonable agreement with
the experimental value $\approx 50$ ms.

   The JILA group has measured the
damping rate of oscillations in the plane perpendicular to the axis of the
trap \cite{jila2}. The trap frequencies in this experiment are $\omega_x =
\omega_y = \omega_z/\sqrt{8} = 2 \pi \nu_{\perp}$, where 
$\nu_{\perp} \approx 129$ Hz. For $N=8 \times 10^4$ at $T=1.1 \,
T_c$ we find that
classically the central density is $\approx 3.7 \times 10^{13}$
cm$^{-3}$.  Assuming that $a = 53$ \AA\, for rubidium atoms
\cite{rbscl}, our calculations of the relaxation times  
Eqs.\,(\ref{simple}), (\ref{half}), (\ref{mz8}) and (\ref{gamma0impp1}) 
imply that the $m_z=0$ mode has a damping time of
$\approx 88$ ms, and the $|m_z|=2$ mode has a damping time of $\approx
29$ ms. The observed damping time was, however, $\approx 50$ ms for the
$m_z=0$ mode, and $\approx 100$ ms for the $|m_z|=2$ mode, indicating
the need for further work to resolve this discrepancy.

\section{Relaxation of thermal anisotropies}
\subsection{Form of the trial function}

    We turn now to temperature relaxation.  As we mentioned in the 
Introduction, in the experiments described in Refs.\,
\cite{Monroe,dalibard,Myatt,myatt,Davis} the energies 
associated with the motion of atoms in the various directions in the 
trap are unequal and to a first 
approximation these anisotropies can be characterized by ascribing
different temperatures to the motions in different directions.
This problem is essentially the same as the one we 
presented earlier.  The most general form of the change in the 
particle distribution function $\delta f_T$ is in this case any 
function $p$ of the energies $E_i$ along the direction $i$,
\begin{eqnarray}
  \delta f_T = p(E_x, E_y, E_z).
\label{changegen}
\end{eqnarray}
It can be checked easily that this form of $\delta f_T$ satisfies
the Boltzmann equation in the absence of collisions with a
frequency equal to zero. 
If one changes the temperature associated with particle motion in the
$z$ direction by an amount $\delta T$ and that associated with motion in
the transverse direction by an amount $-\delta T/2$, conserving in this 
way the average particle energy and the density up to first order 
in $\delta T$,
the change in the particle distribution function $\delta f_T$ is thus
\begin{eqnarray}
   \delta f_T &\propto& f^0(1+f^0) \left[ E_z- \frac 1 2 (E_x+E_y) 
  \right] \nonumber \\
 &\propto& \frac 3 2 f^0(1+f^0) \left[ E_z- \frac E 3 \right].
\label{change}
\end{eqnarray}
Since the results do not depend on the 
anisotropy of the trap, we assume that the external
potential $V$ is isotropic with a frequency $\omega$.
As our initial trial function we therefore take a function 
proportional to $E_z- E/3 = (p_z^2 - p^2/3)/2m + m \omega^2
(z^2 - r^2/3)/2$, 
\begin{eqnarray}
 \Phi_T 
        = p^2 \, P_2(\theta_{\bf p})  + (m \omega r)^2
    \, P_2(\theta_{\bf r}),
\label{phinew}
\end{eqnarray}
where $\theta_{\bf p}$ is the polar angle of the momentum vector,
and $\theta_{\bf r}$ is the polar angle of the position vector.
Later we shall consider improved trial functions.

\subsection{Calculation of the thermal relaxation rate}

   Let us now calculate the thermal relaxation rate $\Gamma_T$. 
For the trial function $\Phi_T$ given by Eq.\,(\ref{phinew})
the only part of it that gives a non-zero contribution when
acted on by the collision integral is the term proportional
to $p^2 \cos^2(\theta_{\bf p}) = p_z^2$.
Since in the previous problem of the damping of oscillation modes
we assumed that $\Phi = (p_z + i m \omega_z z)^2$, the quantity
$\langle \Phi_T^* \,  I[\Phi_T] \rangle$ for thermal relaxation is the same
as that for the relaxation of oscillations, Eq.\,(\ref{nott2}).

  Turning to the denominator $D_T = \langle |\Phi_T|^2 f^0 (1+f^0) \rangle$
which appears in the thermal relaxation rate, the cross terms
that arise from squaring $\Phi_T$ vanish because of the orthogonality of 
the Legendre polynomials in Eq.\,(\ref{phinew}), and therefore
\begin{eqnarray}
     D_T &=& \frac 4 {45} \int d{\bf r} \int d\tau \,
    \left[p^4 + (m \omega r)^4 \right] f^0 (1+f^0)
\nonumber \\
   &=& \frac 8 {45} \,
  \int d{\bf r} \int d\tau \, p^4 f^0 (1+f^0).
\label{denominnew}
\end{eqnarray}
Comparing the above equation with Eq.\,(\ref{denomin2}), $D_T = 
D/3$. Finally we find that 
\begin{eqnarray}
  \Gamma_T \le \frac {\langle \Phi_T^* \, I[\Phi_T] \rangle}
 {\langle |\Phi_T|^2 f^0 (1+f^0) \rangle} = 3 \, \Gamma_0, 
\label{gammatnew}
\end{eqnarray}
or 
\begin{eqnarray}
     \Gamma_T \leq \frac{1}{2} \frac {\int d{\bf r}
    \,\tau_{\eta, {\rm var}}^{-1}
  \langle (p_z^2 - p^2/3)^2 f^0 (1+f^0) \rangle_{\bf p}}
 {\int d{\bf r} \, \langle (p_z^2 - p^2/3)^2 f^0 (1+f^0) \rangle_{\bf p}}.
\label{rate}
\end{eqnarray}
This result may be understood by noting that the decay rate for temperature 
anisotropies in a uniform gas is given by $\Gamma_T \le \tau_{\eta,
{\rm var}}^{-1}$ in the corresponding approximation. 
In a trap there are two facts that must be taken into account. The first 
is that the system is inhomogeneous, and therefore the relaxation time 
must be averaged over space. This accounts for the ratio of integrals in  
Eq.\,(\ref{rate}). The second is that the total energy associated with 
a temperature 
anisotropy has contributions from the potential energy as well as from the 
kinetic energy. For an harmonic trap, the average potential and kinetic 
energies associated with motion along a given axis are equal and therefore the 
total energy for a given direction is twice the kinetic energy. Since 
collisions isotropize directly only the kinetic energy, the time to isotropize 
both the kinetic and potential energies is twice as large as that calculated 
neglecting the effect of the potential energy. This is the origin of the 
factor of $1/2$ in Eq.\,(\ref{rate}).

  Neglecting the effects of quantum degeneracy, Eq.\,(\ref{rate}) gives
\begin{eqnarray}
   \Gamma_T \leq \frac{1}{8 \sqrt 2} n_{\rm cl}(0) \overline{\sigma(u) u}.
\label{ratee}
\end{eqnarray}
Equation (\ref{ratee}) gives for an energy-independent cross
section $\sigma(u)=\sigma_0$,
\begin{eqnarray}
  \Gamma_T \le  \frac 1 {5} n_{\rm cl}(0) \sigma_0 \bar{v}.
\label{temprel3}
\end{eqnarray}
Considering the case of a more general form for the scattering
cross section, Eq.\,(\ref{ratee}) can be written in the form
\begin{eqnarray}
   \Gamma_T \leq
   \frac {1} {15} n_{\rm cl}(0) \sigma_0 \bar{v} \, F(t).
\label{temprel5}
\end{eqnarray}
As defined earlier, $t=T/T_0$, with $k_B T_0 = \hbar^2/m 
a^2$ being the characteristic energy that enters the 
scattering cross section and
\begin{eqnarray}
  F(t)= \int_0^{\infty} dx_r\, x_r^7 e^{-x_r^2} \, \bar{\sigma}.
\label{temprell6}
\end{eqnarray}
For the specific choice $\bar{\sigma}=(1 + x_r^2 t)^{-1}$, the above
expression takes the form
\begin{eqnarray}
  F(t)= \int_0^{\infty} \frac {x_r^7 e^{-x_r^2}\, dx_r} {1 + x_r^2 t}.
\label{temprel6}
\end{eqnarray}
For $T \ll T_0$ the scattering cross section is essentially
energy independent, $\sigma(u) \simeq \sigma_0$, whereas in the opposite
limit the cross section is resonant, $\sigma(u) \simeq 8 \pi/k^2$.
For Cs atoms, a lower bound on the magnitude of their scattering length has
been determined to be $260 \, a_B$
\cite{dalibard}, where $a_B$ is the Bohr radius; therefore $T_0 \le 15$
$\mu$K. Figure 2 shows the result of Eq.\,(\ref{temprel5})
for Cs atoms, for a peak density $n_{\rm cl}(0)= 10^{10}$ cm$^{-3}$.
The $x$ axis shows the temperature in units of $T_0$ and the $y$ axis
shows $\Gamma_T$ measured in units of 
\begin{eqnarray}
   \Gamma_{T,0} = \frac 1 {15} \,
  n_{\rm cl}(0) \sigma_0 \bar{v}_0,
\label{gamaat0}
\end{eqnarray}
where $\bar{v}_0$ is the velocity $\bar{v}_0 = (8 \pi k_B T_0/m)^{1/2}$.
\noindent
\begin{figure}
\begin{center}
\epsfig{file=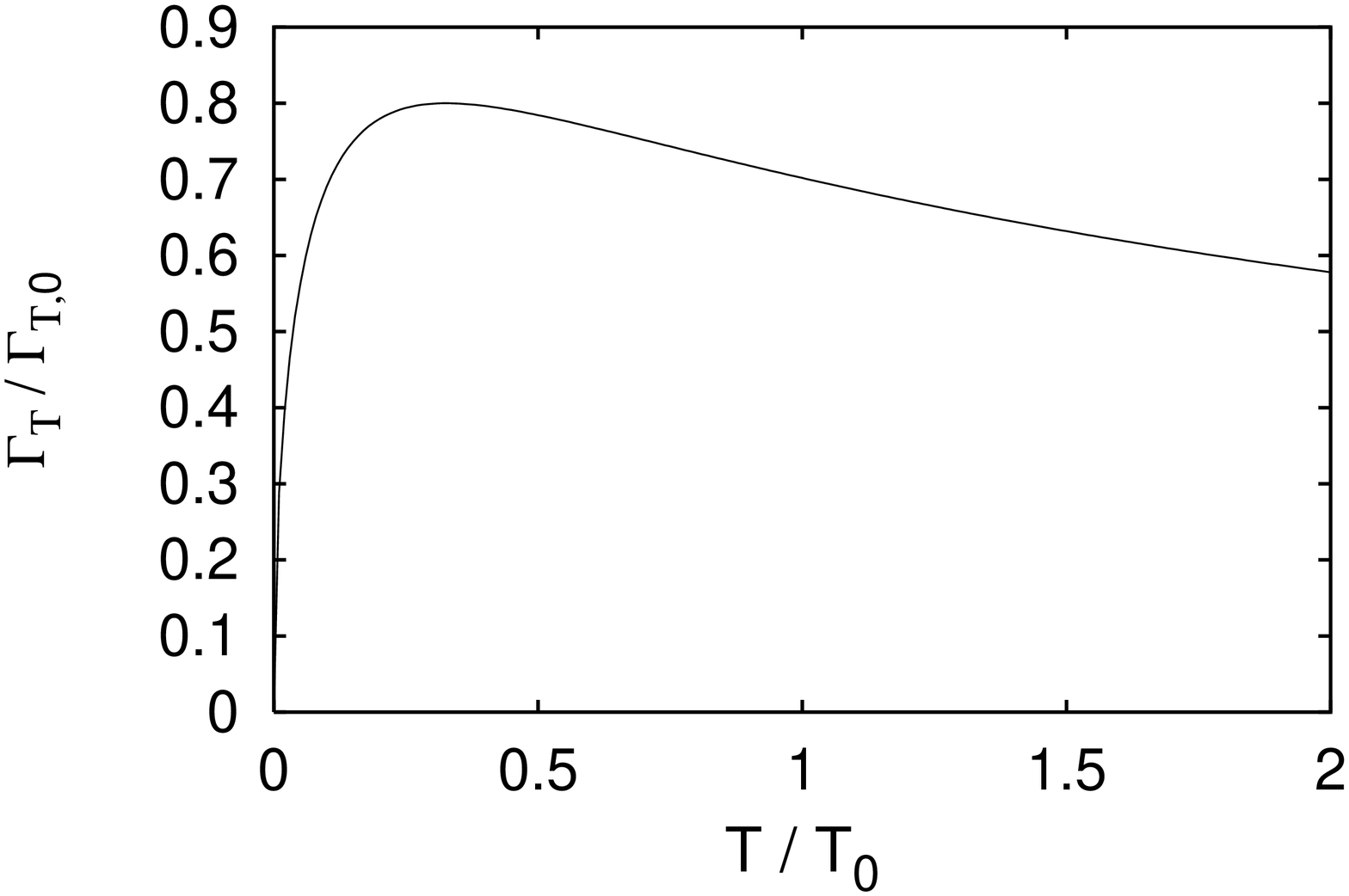,width=7.0cm,height=\linewidth,angle=-90}
\begin{caption}
{The thermal relaxation rate $\Gamma_T$ as a function of $T/T_0$,
as given by the classical version of Eq.\,(\ref{temprel5}),
for a cross section
$\sigma(k) = 8 \pi a^2/[1+(k a)^2]$. 
$\Gamma_T$ is measured in units of $\Gamma_{T,0} = n_{\rm cl}(0) 
\sigma_0 \bar{v}_0/15$.}
\end{caption}
\end{center}
\label{FIG2}
\end{figure}
\noindent
A simple interpolation
formula for the ratio $\Gamma_T/\Gamma_{T,0}$ is given by 
\begin{eqnarray}
    \frac {\Gamma_T} {\Gamma_{T,0}} = \frac {3 \sqrt t} 
   {[1 + (3 t)^{\nu}]^{1/\nu}}, 
\label{interpolation}
\end{eqnarray}
where $\nu \simeq 0.9$.
For Cs atoms we find that $\Gamma_{T,0} \ge 0.2$ s$^{-1}$,
assuming that the magnitude of the scattering length is greater than,
or equal to $260 \, a_B$ \cite{dalibard} and for $n_{\rm cl}(0)= 
10^{10}$ cm$^{-3}$, which is the typical density in the experiments
reported in Ref.\,\cite{dalibard} performed with Cs atoms.

  Equation (\ref{ratee}) for $\Gamma_T$
is useful for the case of Cs atoms; under the experimental conditions
\cite{dalibard} it is a rather good approximation to assume that the
scattering is resonant, with $\sigma = 8 \pi/k^2$, where $k=m
u/2\hbar$ is the relative momentum, since $T > T_0$.
We then obtain the result in the classical regime
\begin{eqnarray}
  \Gamma_T \leq
 \frac {64} {15} \, \frac {n_{\rm cl}(0) \hbar^2}
{m^2 \bar{v}}.
\label{temprel4}
\end{eqnarray}

   It is instructive to define an effective cross section
$\sigma_{\rm eff}$ according to
\begin{eqnarray}
  \sigma_{\rm eff} = \frac {\Gamma_T} {n_{\rm av} \bar{v}}.
\label{seff}
\end{eqnarray}
\noindent
\begin{figure}
\begin{center}
\epsfig{file=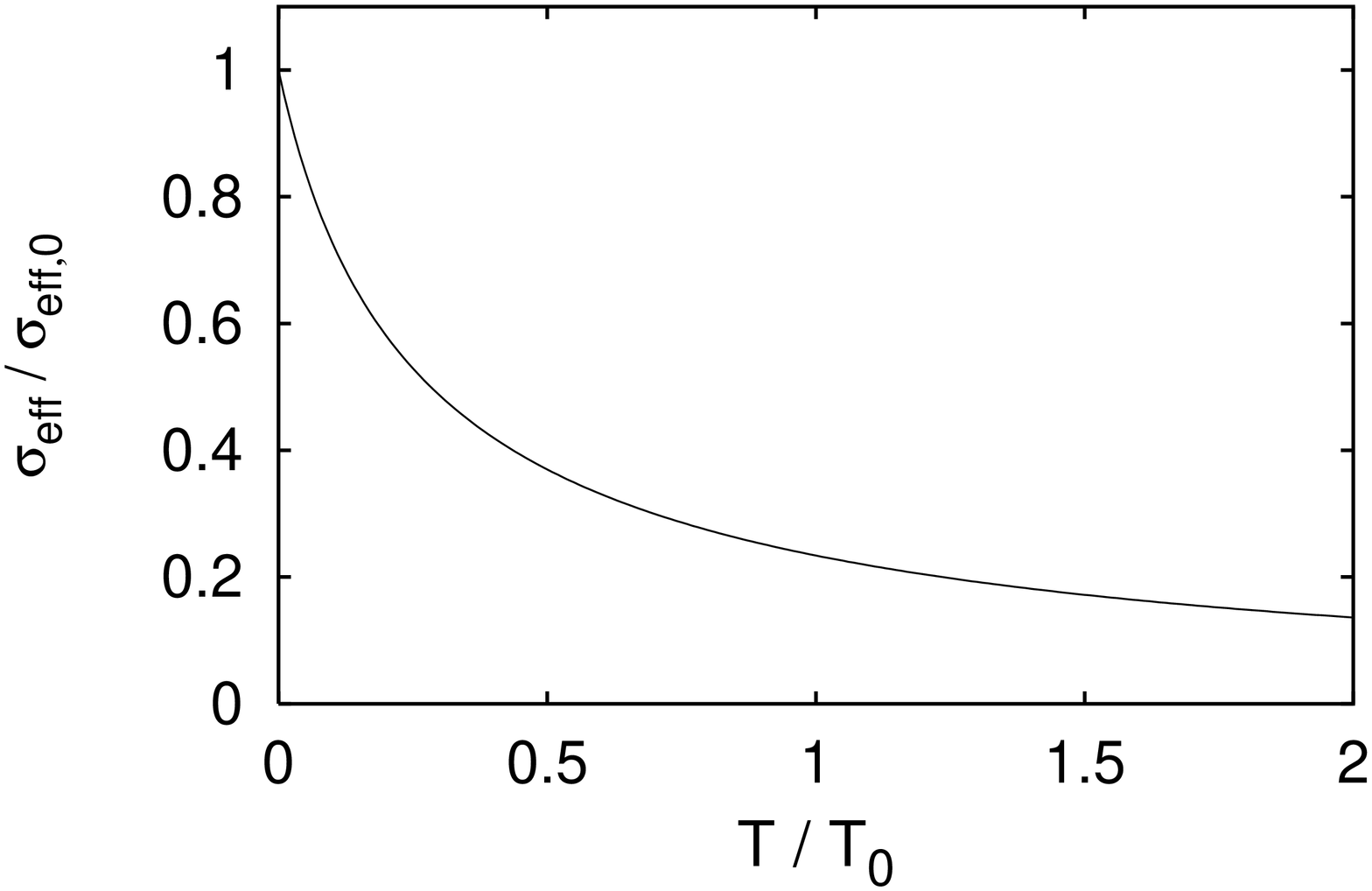,width=7.0cm,height=\linewidth,angle=-90}
\begin{caption}
{The effective cross section $\sigma_{\rm eff}$ given by
Eq.\,(\ref{seff1}),
as a function of $T/T_0$, for a cross section of the form
$\sigma(k) = 8 \pi a^2/[1+(k a)^2]$. The
cross section is measured in units of $\sigma_{\rm eff,0}$
given by Eq.\,(\ref{seff2}).}
\end{caption}
\end{center}
\label{FIG3}
\end{figure}
\noindent
From Eq.\,(\ref{temprel5}) we see that
\begin{eqnarray}
  \sigma_{\rm eff} = \frac {2 \sqrt 2 } {15} \, \sigma_0 F(t).
\label{seff1}
\end{eqnarray}
Figure 3 shows $\sigma_{\rm eff}$ given by Eq.\,(\ref{seff})
as function of $T/T_0$. For energy-independent scattering,
\begin{eqnarray}
  \sigma_{\rm eff} = \frac {2 \sqrt 2} 5 \, \sigma_0 \equiv 
\sigma_{\rm eff,0},
\label{seff2}
\end{eqnarray}
whereas for resonant scattering,
\begin{eqnarray}
   \sigma_{\rm eff} = \frac {128 \sqrt 2} {15} \, \frac {\hbar^2}
 {(m {\bar v})^2}.
\label{seff3}
\end{eqnarray}
Finally using Eq.\,(\ref{seff1}) we plot in Fig.\,4 the effective cross
section $\sigma_{\rm eff}$ for Cs atoms as a function of the temperature
for four different values of the scattering length, $a = 100
\,a_B, 200\,a_B, 300\,a_B$, and $400\,a_B$. For very small values of the
temperature $\sigma_{\rm eff} = \sigma_{\rm eff,0} = 0.142, 0.569,
1.278$ and 2.272 $\times 10^{-11}$ cm$^2$, respectively.
\noindent
\begin{figure}
\begin{center}
\epsfig{file=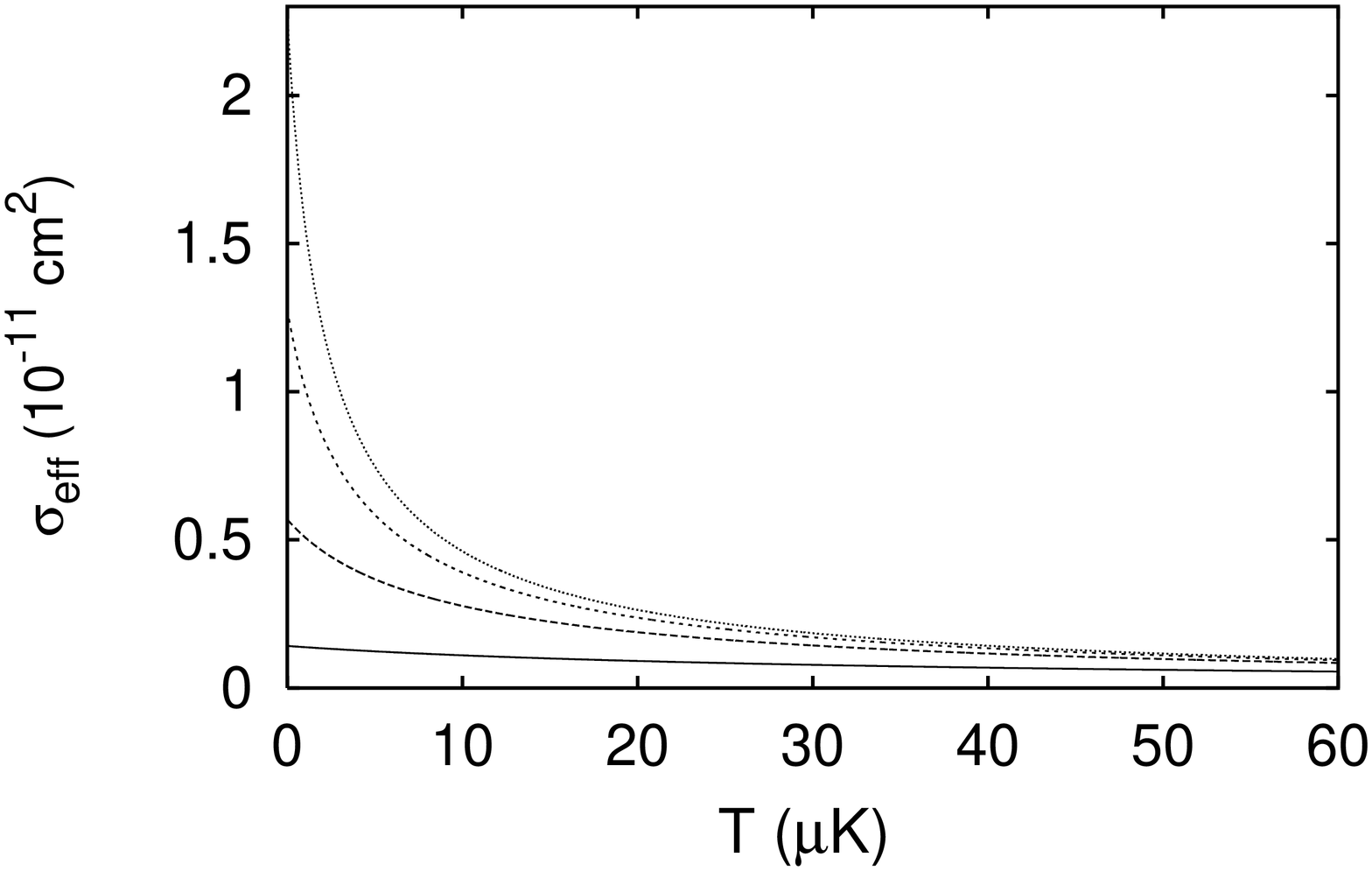,width=7.0cm,height=\linewidth,angle=-90}
\begin{caption}
{The effective cross section $\sigma_{\rm eff}$ for Cs atoms given by
Eq.\,(\ref{seff1}), as a function of $T$, for a cross section of the
form $\sigma(k) = 8 \pi a^2/[1+(k a)^2]$.
The four lines correspond to the following choices for the
scattering length, $a = 100\, a_B, 200 \,a_B, 300 \,a_B$, and
$400 \, a_B$ from bottom to top.}
\end{caption}
\end{center}
\label{FIG4}
\end{figure}

\subsection{Improved estimates}

   Following the same method as in Sec.\,III C we calculate the
corrections to the thermal relaxation rate $\Gamma_T$ using better
trial functions. We choose the improved trial function to be
\begin{eqnarray}
   \Phi_{T,\rm imp} = (1 - \alpha^2)^{1/2} \Phi_{T,0} + \alpha 
\Phi_{T,1},
\label{phinewtimp}
\end{eqnarray}
where $\alpha$ is a variational parameter. $\Phi_{T,0}$ is
given by Eq.\,(\ref{phinew}) and for $\Phi_{T,1}$ we take
\begin{eqnarray}
    \Phi_{T,1} = \sqrt 5 \, 
  \left[ (p_z^2-p^2/3 + (m \omega)^2 (z^2 - r^2/3) \right] \nonumber \\
\phantom{XXXXXXXXX} \times ( 1 -   {E} / {5 k_B T} )
\label{phitnew2exc}
\end{eqnarray}
by analogy with Eq.\,(\ref{measure4}).
In this case we find that only the parts proportional to
$E \, p_z^2 - p^4/6m$ give non-vanishing terms in addition
to the terms we considered earlier with $\Phi_{T,0}$ only, and thus
\begin{eqnarray}
    \Delta[\Phi_{T,\rm imp}] =
     3 \, (p_{r,z}^2 -{p'_{r,z}}^2)
   \left[ (1 - \alpha^2)^{1/2} + \alpha \sqrt 5
\right. \nonumber \\ \left.
 - \frac {\alpha \sqrt 5} 5 \, \frac V {k_B T}
 + \frac {\alpha \sqrt 5} {10} \, \frac 1 { m k_B T} \left( p^2 + \frac 
{P^2} 4
  \right) \right]
 \nonumber \\
- \frac {\alpha \, 3 \sqrt 5} {10} \, \frac 1 {m k_B T} \, 
 (P_z p_{r,z} {\bf P} \cdot {\bf p}_r - P_z p'_{r,z} {\bf P} \cdot {\bf p}'_r)
\nonumber \\
+ \frac {\alpha \sqrt 5} {10} \, \frac 1 {m k_B T} \left[ ({\bf p}
 \cdot {\bf P})^2 -
({\bf p}' \cdot {\bf P})^2 \right].
\label{dphithermal}
\end{eqnarray}
For the case of energy-independent scattering, we find that the matrix 
elements $\gamma_{1,0}/\gamma_{0,0}= -\sqrt 5 /20$, and 
$\gamma_{1,1}/\gamma_{0,0}= 277/240$. Therefore for energy-independent
scattering the thermal relaxation rate
$\Gamma_{T,\rm imp}$ corresponding to $\Phi_{T,\rm imp}$ is given by
\begin{eqnarray}
   \Gamma_{T,\rm imp} (\alpha) = \Gamma_T
\left[ 1 - \frac {\sqrt 5} {10} \alpha (1 - \alpha^2)^{1/2}
 + \frac {37} {240} \alpha^2 \right].
\label{finvaralphath}
\end{eqnarray}
The minimum is at 
$\alpha \approx 0.47$ and the value of the function is $\approx 0.941 \,
\Gamma_T$, i.e., the correction is $\approx 6\%$.
Therefore, our final answer for $\Gamma_{T,\rm imp}$ for the case of 
energy-independent scattering is
\begin{eqnarray}
   \Gamma_{T,\rm imp} \simeq 0.189 \,
  n_{\rm cl}(0) \sigma_0 \bar{v}.
\label{add1}
\end{eqnarray}
In a study of the problem of temperature relaxation, Myatt
\cite{myatt} has used the Boltzmann equation to obtain an analytic
expression for the temperature relaxation rate, and has solved the
problem numerically for large temperature anisotropies.  The
coefficient $1/5$ which appears in Eq.\,(\ref{temprel3}) is consistent
with his results. In a numerical approach of the same problem, Monroe
{\it et al.} \cite{Monroe} have found that a Monte Carlo calculation
gives for this coefficient a value of 0.185, which is remarkably close
to our result, 0.189.

   For the case of resonant scattering we find $\gamma_{1,0}/
\gamma_{0,0}= \sqrt 5 /20$, and $\gamma_{1,1}/\gamma_{0,0}=
53/48$ and thus the correction is $\approx 7\%$ in this case, or
\begin{eqnarray}
   \Gamma_{T,\rm imp} \simeq 4 \,
  \frac {n_{\rm cl}(0) \hbar^2} {m^2 \bar{v}}.
\label{add2}
\end{eqnarray}
As shown by Arndt {\it et al.} in Ref.\,\cite{dalibard},
a Monte Carlo calculation for $\Gamma_T$ for the case of resonant
scattering gives a value for this coefficient of $\approx 64/(2 \times
10.7) \approx 64/21 \approx 3$, instead of the coefficient 4 that
appears in Eq.\,(\ref{add2}).

\subsection{Comparison with experiment}

   The experiment performed by Arndt {\it et al.} \cite{dalibard} with 
Cs atoms shows that the collisional cross section is inversely 
proportional to the temperature in the range $5 - 60$ $\mu$K. 
This indicates that $T_0 \alt 5$ $\mu$K. Since 
these temperatures are larger than $T_0$ for Cs, according to our 
study the cross section should indeed vary as $T^{-1}$. 
Our result given by Eq.\,(\ref{add2}) fits the data of 
Ref.\,\cite{dalibard} rather well.

   In the experiment of Monroe {\it et 
al.} \cite{Monroe} the collisional cross section of Cs atoms was 
measured to be $\approx 1.5 \times 10^{-12}$ cm$^2$, and it was 
independent of the temperature in the range between $\approx 30 - 250$ 
$\mu$K. This value of the cross section is of the same order as
the data in Ref.\,\cite{dalibard} for the highest temperatures.
As Fig.\,4 implies, our calculated cross sections for these temperatures
are also of order $10^{-12}$ cm$^2$.

  In the experiment of Davis {\it et al.} \cite{Davis}, a quadrupole
trap which gives a potential proportional to $(x^2 + y^2 + 4 z^2)^{1/2}$
was used to confine sodium atoms.
The atoms were observed to oscillate 
with an effective frequency of the order of kHz. A non thermal
distribution was created by temporarily displacing the trap center along the 
$z$ axis. The time for relaxation to equilibrium was measured and found
to be of the order of seconds, for a temperature of 200 $\mu$K and for
densities in the center of the trap of order $10^{10}$ cm$^{-3}$.
 
  We cannot immediately apply our results to this experiment because
the potential is not separable in cartesian coordinates. However
Monte Carlo simulations show that motions are effectively separable
(non ergodic) for periods of at least 10 s \cite{Davis}. To estimate the 
relaxation time for temperature anisotropies, we observe that the rate at 
which the particle distribution is isotropized in momentum space by 
collisions is given by the ratio of the integrals in Eq.\,(\ref{rate}).
If the potential were separable in cartesian coorinates and were a
homogeneous function of the coordinates of degree $\nu$, the
potential energy associated with motion in any direction would be
$\nu/2$ times the kinetic energy, and therefore the relaxation
rate allowing for the potential energy would be
\begin{eqnarray}
     \Gamma_{T,\nu} \leq \frac{1}{1+\nu/2} \frac {\int d{\bf r}
    \,\tau_{\eta, {\rm var}}^{-1}
  \langle (p_z^2 - p^2/3)^2 f^0 (1+f^0) \rangle_{\bf p}}
 {\int d{\bf r} \, \langle (p_z^2 - p^2/3)^2 f^0 (1+f^0) \rangle_{\bf p}}.
\label{ratelast2}
\end{eqnarray}
This result reduces to Eq.\,(\ref{rate}) for an harmonic potential ($\nu=2$),
and to the expression $\Gamma_T \le \tau_{\eta, {\rm var}}^{-1}$ 
for a uniform gas ($\nu=0$).
Because the mixing times due to the non-separability of the trapping
potential are long, we would expect that for the quadrupole trap
this result, with $\nu=1$, should give a good estimate of the thermal
relaxation rate. For classical statistics, energy-independent scattering
and $\nu=1$, Eq.\,(\ref{ratelast2}) gives
\begin{eqnarray}
     \Gamma_{T,l} \leq
   \frac 1 {15 \sqrt 2} \, n_{\rm cl}(0) \sigma_0 \bar{v}.
\label{temprellinear1}
\end{eqnarray}
The spatial average that occurs in the present case is 
$\int n_{\rm cl}^2({\bf r}) \, d{\bf r} / \int
n_{\rm cl}({\bf r}) \, d{\bf r} =n_{\rm cl}(0)/2^3$, instead of
$n_{\rm cl}(0)/2^{3/2}$ for harmonic potentials.

  The relaxation rate that was measured in Ref.\,\cite{Davis}
can be written as
\begin{eqnarray}
  \Gamma_{T,l}  \approx
 \frac 1 {4.88} \, n_{\rm cl}(0) \sigma_0 \bar{v},
\label{expmitth}
\end{eqnarray}
if we assume that $a = 27.5$ \AA \, for sodium atoms \cite{Dav}.
Comparing Eqs.\,(\ref{temprellinear1}) and (\ref{expmitth})
we see that there is a difference of a factor $\approx 4.3$. 
We regard the agreement as satisfactory because of the uncertainty
in our estimate due to the unknown role of the lack of separability
of the potential, and because of the difficulty of measuring particle
densities experimentally \cite{private}.

\section{Concluding remarks}

  In this paper we have calculated the damping of low vibrational modes and of
thermal anisotropies using a Boltzmann equation approach.  The main
difference between the present problem and that of calculating the rate of
relaxation processes in a homogeneous gas is that the distribution function
has a more complicated spatial variation.  This is because the motion is
essentially collisionless, and consequently the distribution function is a
function of both momentum and position, whereas for a homogeneous gas in
the hydrodynamic limit the relevant distribution function is a function
only of momentum, apart from a rather simple spatial dependence.  We
formulated the problem as a variational one, and found that for the
simplest trial functions the damping times for vibrational modes and for
thermal relaxation differ only by numerical factors.  With the simplest
variational functions, the most long-lived mode in thermal relaxation
corresponds to an anisotropy of the temperature, as Monroe {\it et al.} assumed
in their calculation.  This picture is not true in general, and an improved
trial function for this mode led to a damping time 6\% longer than the
simplest variational estimate, and in agreement with what was found in
Monte Carlo simulations of the collisional dynamics 
for a constant cross section.  For a
resonant cross section there appears to be a significant discrepancy
between our results and the Monte Carlo simulations of the collisional
dynamics of 
Ref.\,\cite{dalibard}, and the reason for this is at 
present unclear.  For damping of
vibrational modes, the simplest trial function gives a very good
description, and an improved one gave changes in the damping time of less
than one per cent.  If necessary, further improvement of our estimates is
possible by employing more general trial functions.  The existence of
rather precise expressions for relaxation rates should prove a useful tool
in analysis of experimental data.

  Our calculations were carried out under the assumption that the
semiclassical approximation is valid, not only for the streaming terms
in the Boltzmann equation, but also for the collision term.  For smaller
numbers of particles, and/or larger trap frequencies the effects of the
discreteness of the spectrum of single-particle energies in the trap will
become important, and the situation will be closer to that encountered in
studies of collective motion in finite nuclei \cite{broglia}.

\vskip1pc

\begin{appendix}

\section{Quantum degeneracy corrections}

    Including effects of quantum degeneracy to leading order in the
fugacity $z({\bf r})$, we find for energy-independent scattering,
$\sigma(u) = \sigma_0$, that the collision integral in Eq.\,(\ref{corr0})
gives
\begin{eqnarray}
   \bar{I_c} [z({\bf r})] = z^2 \, \sqrt \pi
  \left( 1 + z \frac 9 {16} \sqrt {\frac 3 2} \right).
\label{expnum}
\end{eqnarray}
Expanding the denominator of Eq.\,(\ref{corr0}) we find
\begin{eqnarray}
    \bar{D}({\bf r}) = z \, \frac {3 \sqrt \pi} 8
  \left( 1 + z \frac 1 {4 \sqrt 2} \right).
\label{expden}
\end{eqnarray}
Combining Eqs.\,(\ref{expnum}) and (\ref{expden}) we thus find
that Eq.\,(\ref{corr0}) takes the form
\begin{eqnarray}
     \Gamma_0 \leq \frac{4}{15 \sqrt 2} \,
   \left( \frac {m k_B T} {2 \pi \hbar^2} \right)^{3/2} \, \sigma_0 \,
\bar{v} \times
\phantom{XXXX}
\nonumber
\end{eqnarray}
\begin{equation}
    \int z^2({\bf r})
  \left(1 + z({\bf r}) \frac {9} {16} \sqrt {\frac 3 2}
 \right) \, d{\bf r}
\left/
   \int z({\bf r}) \left(1 +  \frac {z({\bf r})}
  {4 \sqrt{2}} \right)
 \, d{\bf r} \right..
\label{corr1}
\end{equation}
Performing the spatial integrations in Eq.\,(\ref{corr1}), we
find for a harmonic potential
\begin{eqnarray}
     \Gamma_0 \leq \frac{1}{15} \, \sigma_0 \bar{v}
   \left( \frac {m k_B T} {2 \pi \hbar^2} \right)^{3/2}
  z(0) \left[ 1 + \frac {5 z(0)} {16} \right],
\label{corr2}
\end{eqnarray}
where $z(0)$ is the fugacity calculated at the center of the cloud,
${\bf r}=0$. We now calculate $z(0)$ as a function of the
total number of particles
$N = \sum_{\bf p} \int d{\bf r} \, f^0({\bf r},{\bf p})$, which gives
up to order $z^2(0)$
\begin{eqnarray}
 N = \left( \frac {k_B T} {\hbar \bar{\omega}} \right)^3 z(0)
  \left( 1 + \frac {z(0)} 8 \right).
\label{corr3}
\end{eqnarray}
Solving the above equation in terms of $z(0)$ up to $N^2$ and
using the expression for the critical temperature
$k_B T_c = \hbar \bar{\omega} [N/\zeta(3)]^{1/3}$, we find
\begin{eqnarray}
   z(0) = \left( \frac {T_c} T \right)^3 \zeta(3)
  \left[ 1 - \frac {\zeta(3)} 8 \left( \frac {T_c} T \right)^3 \right].
\label{corr4}
\end{eqnarray}
Inserting Eq.\,(\ref{corr4}) into Eq.\,(\ref{corr1}) we obtain
\begin{eqnarray}
    \Gamma_0 \leq \frac{1}{15} \, n_{\rm cl}(0)
  \sigma_0 \bar{v}
 \left[ 1 + \frac 3 {16} \zeta(3) \left( \frac {T_c} T \right)^3
\right],
\label{appcorrnew}
\end{eqnarray}
where $n_{\rm cl}(0) = N  \bar{\omega}^3
[m /2 \pi k_B T]^{3/2}$.

\end{appendix}

\vskip1pc

   G.M.K. was supported by the European Commission, TMR program,
contract No.\,ERBFMBICT 983142. Helpful discussions with G. Baym,
P. Cvitanovi\'{c}, W. Ketterle, D. G. Ravenhall and C. E. Wieman are
gratefully acknowledged. We should also like to thank C. E. Wieman
for providing us with a copy of Ref.\,\cite{myatt}. G.M.K. would
like to thank the Foundation of Research and Technology, Hellas
(FORTH) for its hospitality.


\vspace{1cm}



\begin{references}

   \bibitem{jila1} D. S. Jin, J. R. Ensher, M. R. Matthews,
C. E. Wieman, and E. A. Cornell, Phys. Rev. Lett. {\bf 77}, 420 (1996).

    \bibitem{jila2} D. S. Jin, M. R. Matthews, J. R. Ensher, C. E. 
Wieman, and E. A. Cornell, Phys. Rev. Lett. {\bf 78}, 764 (1997).

    \bibitem{mit} M.-O. Mewes, M. R. Andrews, N. J. van Druten,
D. S. Durfee, C. G. Townsend, and W. Ketterle, Phys. Rev. Lett. {\bf 77},
988 (1996).

    \bibitem{Ketnew}  D. M. Stamper-Kurn, H.-J. Miesner, S. Inouye,
M. R. Andrews, and W. Ketterle, Phys. Rev. Lett. {\bf 81}, 500 (1998).

   \bibitem{Monroe} C. R. Monroe, E. A. Cornell, C. A. Sackett, C. J. 
Myatt, and C. E. Wieman, Phys. Rev. Lett. {\bf 70}, 414 (1993).

   \bibitem{dalibard} M. Arndt, M. B. Dahan, D. Gu\'ery-Odelin, M. W.
Reynolds, and J. Dalibard, Phys. Rev. Lett. {\bf 79}, 625 (1997).

   \bibitem{Myatt}  N. R. Newbury, C. J. Myatt, and C. E. Wieman,
Phys. Rev. A {\bf 51}, R2680 (1995).

    \bibitem{myatt}  C. J. Myatt, Ph. D. thesis, Univ. of Colorado
(1997).

   \bibitem{Davis} K. B. Davis, M.-O. Mewes, M. A. Joffe, M. R. Andrews,
and W. Ketterle, Phys. Rev. Lett. {\bf 74}, 5202 (1995).

   \bibitem{kps} G. M. Kavoulakis, C. J. Pethick, and H. Smith,
Phys. Rev. Lett. {\bf 81}, 4036 (1998).

    \bibitem{Wu} H. Wu and C. Foot, J. Phys. B {\bf 29}, L321 (1996).

  \bibitem{Henrik} H. Smith and H. H. Jensen, {\it Transport Phenomena}
(Oxford University Press, Oxford, 1989).

  \bibitem{Landau2} L. D. Landau and E. M. Lifshitz, {\it Fluid Mechanics}
(Pergamon, Oxford, 1959), Sec.\,77.

   \bibitem{Dav} E. Tiesinga, C. J. Williams, P. S. Julienne, K. M. Jones,
P. D. Lett, and W. D. Phillips, J. Res. Natl. Inst. Stand.
Technol. {\bf 101}, 505 (1996).

  \bibitem{Landau} L. D. Landau and E. M. Lifshitz, {\it Quantum 
Mechanics} (Pergamon Press, Oxford, 1977), p. 548.
 
   \bibitem{Griffin} T. Nikuni and A. Griffin, J. Low Temp. Phys. {\bf 111}, 
793 (1998).

   \bibitem{Burnett} D. Burnett, Proc. Lond. Math. Soc. {\bf 39}, 385 (1935).

   \bibitem{landau2} E. M. Lifshitz and L. P. Pitaevskii, {\it
Physical Kinetics} (Pergamon Press, Oxford, 1981), p. 31.

  \bibitem{private} W. Ketterle (private communication).

    \bibitem{hy} A. Griffin, W. C. Wu, and S. Stringari, Phys. Rev. 
Lett. {\bf 78}, 1838 (1997).

   \bibitem{KPS} G. M. Kavoulakis, C. J. Pethick, and H. Smith,
Phys. Rev. A {\bf 57}, 2938 (1998).

   \bibitem{Usama} D. Gu\'ery-Odelin, F. Zambelli, J. Dalibard, and
S. Stringari, e-print cond-mat/9904409; U. Al Khawaja, C. J. Pethick 
and H. Smith, e-print cond-mat/9908043.

    \bibitem{rbscl} J. R. Gardner, R. A. Cline, J. D. Miller,
H. M. J. M. Boesten, and B. J. Verhaar, Phys. Rev. Lett.
{\bf 74}, 3764 (1995); H. M. J. M. Boesten, C. C. Tsai, J. R. Gardner,
D. J. Heinzen, and B. J. Verhaar, Phys. Rev. A {\bf 55}, 636 (1997).

   \bibitem{broglia}  G. F. Bertsch, P. F. Bortignon, and R. A. Broglia,
Rev. Mod. Phys. {\bf 55}, 287 (1983).

\end{references}
\end{document}